\documentstyle[aps,amssymb,psfig,twocolumn]{revtex}  
\begin{document}
\textheight 230mm
\wideabs
{
\title{\LARGE 
{\sf Predicting Crystal Structures with Data Mining of Quantum Calculations}}
\author{
Stefano Curtarolo$^{1}$, Dane Morgan$^{1}$, Kristin Persson$^{1}$, John Rodgers$^{2}$, Gerbrand Ceder$^{1,3}$}
\address{
$^1$Department of Materials Science and Engineering, MIT, Cambridge, MA 02139 \\
$^2$Toth Information Systems Inc., Ottawa, Canada\\
$^3${corresponding author, e-mail: gceder@mit.edu}
}
\date{\today}
\maketitle
\begin{abstract}
Predicting and characterizing the crystal structure of materials is a key
problem in materials research and development.  It is typically addressed
with highly accurate quantum mechanical computations on a small set of
candidate structures, or with empirical rules that have been extracted from
a large amount of experimental information, but have limited predictive
power.  In this letter, we transfer the concept of heuristic rule
extraction to a large library of {\it ab-initio} calculated information, and
demonstrate that this can be developed into a tool
for crystal structure prediction.
\end{abstract}
%\pacs{PACS numbers: 61.50.Ah ?? double check}
}

{\it Ab-initio} methods, which predict materials properties from the fundamental
equations of quantum mechanics, are becoming a ubiquitous tool for
physicists, chemists, and materials scientists. These methods allow
scientists to evaluate and pre-screen new materials ``in silico'', rather
than through time-consuming experimentation, and in some cases, even make
suggestions for new and better materials \cite{ref1,ref2,ref3,ref4,ref5}. One inherent
limitation of most {\it ab-initio} approaches is that they do not
make explicit use of results of previous calculations when studying a new system.
This can be contrasted with data-centered methods, which mine existing data libraries
to help understand new situations. The contrast between data-centered and traditional
{\it ab-initio} methods can be seen clearly in the different approaches used to
predict the crystal structure of materials. This is a difficult but
important problem that forms the basis for any rational materials design.
In heuristic models, a large amount of experimental observations are used in
order to extract rules which rationalize crystal structure with a few
simple physical parameters such as atomic radii, electronegativities, etc..
The Miedema rules for predicting compound forming \cite{ref6}, or the Pettifor maps
\cite{ref7} which can be used to predict the structure of a new binary material by
correlating the position of its elements in the periodic table to those of
systems for which the stable crystal structure is known, are excellent
examples of this. {\it Ab-initio} methods differ from these data-centered methods 
in that they do not use historic and cumulative information about previously studied
systems, but rather try to determine structure by optimizing from scratch
the complex quantum mechanical description of the system, either directly
(as in {\it ab-initio} Molecular Dynamics), or in coarse-grained form (as in
lattice models \cite{ref8a,ref8b,ref8c}). Here, we merge the ideas of 
data-centered methods with the predictive power of {\it ab-initio} computation. 
We propose a new approach whereby
{\it ab-initio} investigations on new systems are informed with {\it knowledge}
obtained from results already collected on other systems.
We refer to this approach as Data Mining of Quantum
Calculations (DMQC), and demonstrate its efficiency in increasing the speed of
predicting the crystal structure of new and unknown materials. 
Using a Principal Component Analysis on over 6000 {\it ab-initio} energy calculations, we
show that the energies of different crystal structures in binary alloys are
strongly correlated between different chemical systems, and demonstrate how
this correlation can be used to accelerate the prediction of new systems.
We believe that this is an interesting new direction to address in a
practical manner the problem of predicting the structure of materials.

Using Density Functional Theory we have calculated a {\it library} of {\it ab-initio}
energies for 114 different crystal structures in each of 55 binary metallic
alloys. About 1/3 of the crystal structures in the library were chosen from
the most common binary crystal structures in the CRYSTMET database for
intermetallics  \cite{ref9}. The rest are superstructures of the fcc, bcc, and hcp
lattices. The alloys include all 45 binaries that can be made from row 4
transition metals, as well as ScAl, AgMg, and 8 binary Ti alloys 
(AgTi, CdTi, MoTi, PdTi, RhTi, RuTi, TcTi, TiZr).
The formation energy for each structure is determined with respect to the most
stable structure of the pure elements. Energy calculations were done using
density functional theory, in the local density approximation, with the
Ceperley-Alder form for the correlation energy as parameterized by Perdew-Zunger \cite{ref10} 
with ultrasoft pseudopotentials, as implemented in VASP \cite{ref11}.
Calculations are at zero temperature and pressure, and without zero-point
motion. The energy cutoffs in an alloy was set to 1.5 times the larger
of the suggested energy cutoffs of the pseudopotentials of the elements 
of the alloy (suggested energy cutoffs are derived by the method described in \cite{ref11}).
Brillouin zone integrations were done using 2000/(number of atoms in unit cell) 
${\bf k}$-points distributed as uniformly as possible on a Monkhorst-Pack mesh.
We verified that with these energy cutoffs and k-points mesh the absolute energy
is converged to better then 10 meV/atom. Energy differences between structures
are expected to be converged to much smaller tolerances.
Spin polarization was not used as no magnetic alloys were studied. 
All structures were fully relaxed.

For each alloy $i$, consider the 114 structural formation energies as the
components of a vector ${\bf E}_i$ in a 114-dimensional space. If the energies of
the structures are linearly dependent then the vectors for each alloy will
not be distributed randomly in the 114 dimensional space, but confined in a
subspace of reduced dimension. To look for such approximate linear
dependencies we use Principal Component Analysis \cite{ref12} (PCA). 
This allow us to express the
energy vector of an alloy as an expansion in a basis of reduced dimension
$d$, ${\bf E}_i=\sum_{j=1}^{d} \alpha_{ij}{\bf e}_i + {\bf \epsilon}_i(d)$, 
where ${\bf \epsilon}_i(d)$ is the error vector for the alloy $i$. PCA consists of
finding the proper basis set $\left\{ {\bf e}_j(d) \right\}$ that minimizes the remaining squared
error $\sum_i  {\bf \epsilon}_i^T  \cdot {\bf \epsilon}_i$ for a given dimension $d$. 
These optimum basis vectors $\left\{ {\bf e}_j(d) \right\}$ are
called the Principal Components (PC's), and form a set of orthogonal
vectors ordered by the amount of variation of the original sample they can
explain. More intuitively, they are a new set of axes in the 114
dimensional space, ordered according to the fraction of the data lying
along that axis. As an extreme example, if the energies of all 55 alloys
were proportional each other, then all the alloy vectors would lie along a
single line, and the first PC would be a subspace that encompassed all the
data $(d = 1)$.

\vspace{-2mm}
\begin{figure}
  \psfig{file=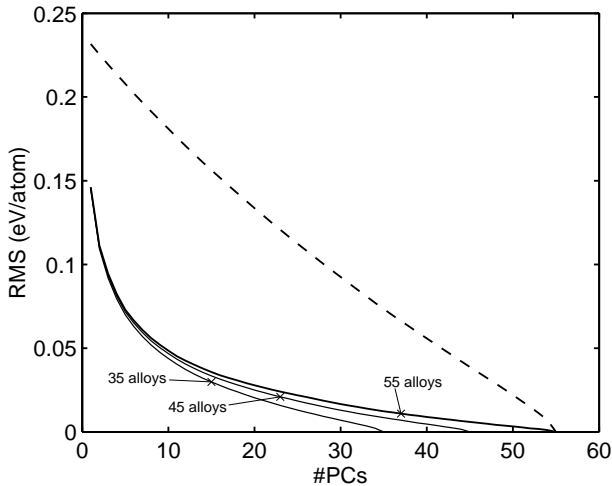,width=82mm,clip=}
  \vspace{2mm}
  \caption{The RMS error as a function of the number of principal components.
    The solid lines show results for the libraries containing 35, 45, and 55
    alloys. The dashed line shows results for the 55-alloy library where the
    energies for each alloy have been randomly permuted.}
  \label{label.fig.1}
\end{figure}
\vspace{-1mm}

A Principal Components Analysis of our {\it ab-initio} data set (Figure \ref{label.fig.1}) shows
that significant dimension reduction is possible in the space of structural
energies. The solid curve, labelled ``55'', in Figure \ref{label.fig.1} shows the remaining
unexplained Root Mean Square (RMS) error (average error in the 114
structural energies of the 55 alloys), as function of the number of
principal components $d$. All quantities are given as energy per atom. The number
of relevant dimensions depends on the error one can tolerate. For example,
to describe the energies with a 50 meV RMS error, only 9 dimensions are
required, much less than the original 114. The implication is that it is
possible to perform far fewer than 114 calculations to parameterize the 9
dimensional subspace, and then derive the other 105 energies through linear
relationships given by the PCA.

Dimension reduction holds only because the energy differences of structures
are strongly correlated between alloys with different chemistry. In fact,
if we perform a PCA analysis in which the structural energies for each
alloy are randomly permuted, and hence destroy their relations, there is
little opportunity for dimension reduction, as the dashed curve in Figure \ref{label.fig.1}
shows. Given an acceptable accuracy, dimension reduction does not
depend on the dimension of the library, once the library is bigger than a
certain size. Figure \ref{label.fig.1} shows the PCA analysis for 35, 45, and 55 alloys.
For subspaces defined by up to $\approx20$ PC's (27 meV RMS accuracy) the variance
is essentially independent of the number of alloys, indicating that the
dimension reduction we obtain can be expected to apply to new alloy systems.

These correlations are further confirmation that the success of heuristic
methods is not accidental, and that with relatively few parameters it can
be possible to predict the structure of a binary alloy. In fact, these
correlations can be used to develop an {\it ab-initio-data-mining} algorithm
that rapidly searches through the available space of possible structures.

Given a library of $N_a$ alloys, $N_s$ structures, and a new alloy where the 
first $n$ energies have been calculated, we predict the energy for structure 
$i > n$ of the new alloy as follows.  
Define ${\bf X}$ as the $(n,N_a)$ matrix of energies for structures $\{1 \dots n\}$ in the library.
Define ${\bf y}$ as the $n-$component vector of known energies for the new alloy and 
${\bf X'}$ as the $N_a$ component vector of energies for structure $i$ for all alloys in the library. 
The scalar ${y'}$ represents the unknown energy of structure $i$ for the new alloy. 
We regress ${\bf y}$ on ${\bf X}$ using the Partial Least Squares method \cite{ref13,ref14} 
implemented with the SIMPLS algorithm\cite{ref15}.  
The resulting regression coefficients are used to predict ${y'}$ from ${\bf X'}$.  
This is done for every structure of the new alloy for which the energy has not yet been calculated. 

The ground states for an alloy are found through an iterative scheme.  
At each step, the PLS regression is used to find the most probable ground state, 
which is then calculated with quantum mechanics and added to the data.  
The algorithm is started with only the pure element energies for the two elements of 
the alloy in the bcc, fcc, and hcp structures, and then proceeds as follows.  \\
{\bf Step 1 (prediction).} The regression algorithm given above is used to predict all unknown structural 
energies in the new system.  We found that for early iterations ($<10$) the RMS 
error can be reduced by preclustering the library into ordering and phase-separating 
systems and regressing only within the library subcluster in which the system is 
predicted to fall. Physically, this means that for early stage of the iterative
 procedure, new alloys regress better with similar alloys than with the complete library.  \\
{\bf Step 2 (suggestion).} 
With the available calculated energies we determine the ground-state energy 
versus composition curve (convex-hull). The convex hull, which is the set of tie lines 
that connects the lowest energy ordered phases, represents the free energy of the
alloy at 0K. Structures with energy above a tie line, 
are unstable with respect to mixtures of the two structures that define the 
vertices of the tie line. Hence, the convex hull determines the phase stability of the 
system at zero temperature. 
The structure with predicted energy farthest below the convex hull of calculated 
energies is calculated with quantum mechanics and added to the database. If no 
structure breaks the hull, we look for the structure predicted to be closest to the hull. 
For early iterations 
($<13$ in Figure 2), if no such structure can be found within 80 meV of the ground 
state hull, we consider the prediction to have failed in this step, and instead add 
the most frequent and not yet calculated ground state structure of the database.  \\
{\bf Step 3 (calculation).} The candidate suggested structure is then {\it calculated} 
with Quantum Mechanics and added to the list, and the entire process is iterated 
({\it prediction $\Rightarrow$ suggestion $\Rightarrow$ calculation}). 
With each step, more energetic information for the new alloy is incorporated and a 
better prediction of the ground state can be expected.

\vspace{-2mm}
\begin{figure}
  \psfig{file=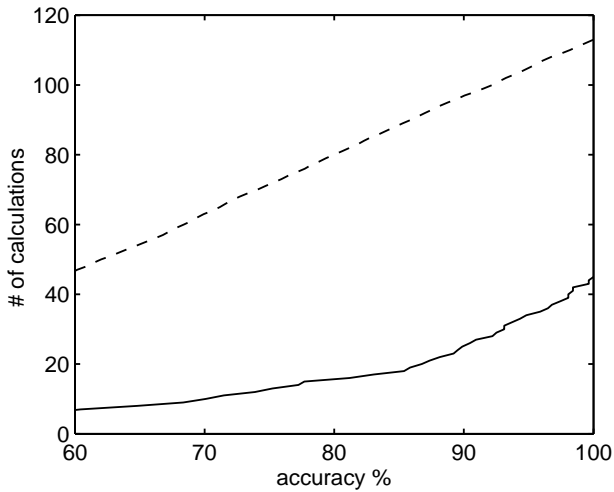,width=82mm,clip=}
  \vspace{2mm}
  \caption{Number of calculations as a function of the percentage of ground
    states predicted correctly, with the DMQC method (solid line) and with
    random structure selection (dashed line). Ninety percent accuracy can be
    achieved with DMQC with 26 calculations, much less than the 98 calculations
    necessary for random structure selection.}
  \label{label.fig.2}
\end{figure}
\vspace{-1mm}

Any structure in the library can be predicted and there are no preconceived
biases as to the symmetry or underlying superlattice of the structure as is
the case for methods that work with lattice model approaches. For example,
in the Ti-Pt alloy, our method correctly finds the A15 \cite{ref16,ref17}  structure to
be a ground state for Ti$_3$Pt after only 20 steps in the algorithm, even
though this structure is not a superstructure of fcc (the structure of Pt)
or hcp (the structure of Ti), and is therefore not an obvious structure 
to investigate for this system.
To study in a more statistically significant way how this iterative scheme
converges we tested how well the library minus one alloy can perform
predictions on the alloy left out. A key property is whether the alloy is
immiscible (no ordered compounds) or has intermediate compounds 
(compound-forming). Empirical schemes like the one developed by Miedema \cite{ref6}  have been
particularly successful in classifying this difference. We find that DMQC
can predict whether an alloy excluded from the library is compound-forming
with 95\%, 98\%, and 100\% accuracy using 3, 6, and 13 calculations,
respectively. Note that here and below, we do not count the initial pure
element calculations, since these are only performed once for each element.
For comparison, if one randomly picked trial structures from the list of
114 structures, predictions with 95\%, 98\%, and 100\% accuracy require 7, 21,
and 98 calculations, respectively. The DMQC method performs extremely well,
far better than a naive random choice of structures, and gives almost
perfect prediction with a small amount of computation.

A more stringent evaluation is whether the correct stable crystal structures are
predicted for the system left out. Figure \ref{label.fig.2} (solid line) shows the number
of calculations required as a function of the percentage of ground states
predicted correctly (averaged over all alloys). For our purpose, ``correct''
is what would be obtained from the direct quantum mechanical calculations
on all 114 structures. Ninety percent accuracy can be achieved with less
than 26 calculations for an alloy. To achieve the same confidence level
with random structure selection (dashed line) one needs to calculate almost
the complete database (98 calculations).

Even though it is generally believed that the binary alloys are well
characterized experimentally, our approach can be used to quickly predict previously
unknown stable structures in some systems. For example, with only 26 calculations we
predict Ag$_3$Cd and Ag$_2$Cd respectively to have the DO$_{24}$ and C37 structure. 
In addition, we predict the previously unidentified
structure for CdZr$_3$ to be A15 (Cr$_3$Si-type).  This prediction takes
only 21 iterations and is particularly interesting since
A15 does not share the hcp parent lattice of Cd and Zr.
These predictions were confirmed by calculation of 
all the structures in the library. A more detailed analysis of the predictions made 
from our database in a large number of systems will be published elsewhere.

\begin{figure}
  \psfig{file=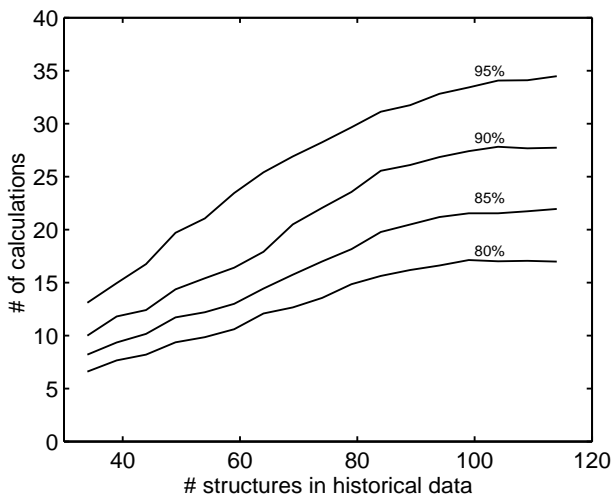,width=82mm,clip=}
  \vspace{2mm}
  \caption{The average number of calculations needed to obtain a given
    accuracy of predicted crystal structures, as a function of the number of
    structures in the library. Results are given for 80\%, 85\%, 90\%, and 95\%
    accuracies. The number of calculations increases less than linearly with
    the number of structures in the database, demonstrating that efficiency
    increases as the library grows.}
  \label{label.fig.3}
\end{figure}

More structures will need to be added to the library to give the method
better applicability to many unknown systems. It is therefore important to
assess how the number of required calculations scales with the number of
structures in the library. This scaling is shown in Figure \ref{label.fig.3} for various
required confidence levels. As the library grows, more calculations are
needed to select between the increasing number of possibilities.
Fortunately, the number of calculations increases less than linearly with
the number of structures in the database, demonstrating that efficiency
increases as the library grows.

Our current DMQC approach has the limitation that structure types must
already be in the database to be predicted. However a concerted effort to
develop a public database, analogous to those used in biology, may make
this limitation less important. Our work has also focussed on a simple test
library of binary alloys. The real payoff will come with the inclusion of
multicomponent systems, where fewer than 10\% of all intermetallic systems
have been characterized \cite{ref7,ref18}. A library of ternary structures can be
integrated with the binary libraries and extensions of the formalism are
not required, besides adding an extra composition variable. Although the
data mining methods discussed here are centered around dimension reduction
and linear correlation approaches, including nonlinear methods (e.g.,
neural nets, clustering, learning machines, etc.) will certainly be more
effective in extracting information from the library. For example, problems
associated with using a single set of regression coefficients for the whole
heterogeneous data set can be avoided by preclustering the data and using
linear regression only within each cluster.

\vspace{-2mm}
\vspace{-1mm}

In summary, by data mining quantum mechanical calculations (DMQC) we have
established that there exist significant correlations among {\it ab-initio}
energies of different structures in different materials. The correlations
we found can be seen as a formal extension of the heuristic structure-properties 
selection rules that have been established in the past on the
basis of large amounts of {\it experimental} structure information \cite{ref18,ref19,ref20}.
Our approach differs from the previous classifications in that we correlate
on calculated information (structural energies in our particular example),
and hence our description can be used when there is limited experimental
data, and can be extended to arbitrary accuracy.

The data-mined correlations form the basis for an efficient algorithm for
structure prediction which has all the capacities of {\it ab-initio} energy
methods, but extracts information from previous calculations on other
systems in order to efficiently propose candidate structures. Because
structures are not found through optimization of some physical variable
space (e.g. atomic coordinates), it has none of the problems with time-scale 
and equilibration common to other approaches. We believe that the
integration of data mining techniques with {\it ab-initio} methods is a promising
development towards the practical prediction of crystal structure.

The research was supported by the Department of Energy, Office of Basic Energy Science
under Contract No. DE-FG02-96ER45571.

\end{document}